# Lattice rotation vortex at the monoclinic ferroelectric domain boundary in relaxor ferroelectric crystal

*Yu-Tsun Shao[1] and Jian-Min Zuo[1, 2*]*

[1]*Department of Materials Science and Engineering, University of Illinois at Urbana-Champaign, Urbana, Illinois 61801, USA*

[2]*Frederick Seitz Materials Research Laboratory, University of Illinois at Urbana-Champaign, Urbana, Illinois 61801, USA*

**Abstract.**

We present the evidence of lattice rotation vortices having an average radius of ~7 nm at the monoclinic ferroelectric domain boundary of $(1-x)Pb(Zn_{1/3}Nb_{2/3})O_3-xPbTiO_3$ (PZN-xPT, x=0.08). Maps of crystal orientations, domain configurations, symmetry breaking are obtained using scanning convergent beam electron diffraction (SCBED). Such measurements suggest the merging of 2D and 1D topological defects, with implications for domain-switching mechanisms in relaxor ferroelectric crystals, and the possibility of a new form of nanoscale ferroelectric devices.

* Corresponding author: jianzuo@illinois.edu

Ferroelectric domain walls (DWs) are topological defects that involve changes in the polarization direction accompanied by small lattice distortions. Having a large density of mobile DWs facilitates domain switching and therefore dramatically enhances the susceptibility of ferroelectrics and piezoelectric coupling coefficients [1]. DWs also exhibit emergent physical properties. For example, charged DWs of $BiFeO_3$ [2] and $Pb(Zr, Ti)O_3$ [3] exhibit a significant conductivity increase compared to bulk materials, which can be advantageous for device applications [4]. The structural determination of DWs requires 1) identifying two neighboring polarization domains, 2) determining the transition structure between the domains, and 3) identifying the nature of the polarization in the transition region.

The ferroelectric domains can be identified by optical microscopy [5], scanning probe microscopy [6], x-ray diffraction [7-10], neutron diffraction [11-15], or electron diffraction techniques [16-18]. Ferroelectric DWs can be categorized by the dipole transition behaviors across the boundary, which are non-chiral DWs (Ising-like), chiral DWs (Bloch- or Neel-like), or mixtures of both [19]. The structural determination of DWs was demonstrated by Nelson *et al.* and Jia *et al.* using atomic resolution electron imaging for tetragonal $Pb(Zr,Ti)O_3$ [20, 21] and rhombohedral $BiFeO_3$ [22] crystals by quantifying the displacements of atomic columns. However, such techniques have yet to be applied to relaxor ferroelectric crystals with monoclinic symmetry.

In lead-based complex perovskite oxides having the chemical formula $(1-x)Pb(B'^{+2}_{1/3}B''^{+5}_{2/3})O_{3-x}-xPb(B'''^{+4})O_3$ (*B'*, *B''*, *B'''*=Zn, Nb, Ti for PZN-PT, and Mg, Nb, Ti for PMN-PT), exceptional piezoelectric properties [23] are obtained at the morphotropic phase boundary (MPB), where nanometer-sized monoclinic domains have been reported by X-ray diffraction [5, 24], neutron diffraction [11-15], and electron microscopy [16, 18, 25]. However, we know very little about the structure of DWs in relaxor ferroelectric crystals and the properties of DWs, because of the experimental challenge of determining monoclinic, and nanometer-sized, domains. Moreover, Houchmandzadeh and coworkers showed that the coupling between two order parameters, such as electric dipoles and strain, can induce chirality at the DWs [26].

Here, we describe a scanning convergent beam electron diffraction (SCBED) study of the DWs in the relaxor-based ferroelectric crystal of PZN-8%PT. Using energy-filtered (EF) SCBED. We have identified nm-sized domains having monoclinic (M) *Pm* symmetry in single crystal PZN-8%PT. A careful examination of the DWs revealed the presence of lattice rotation vortices near DWs. These vortices involve continuous lattice rotation across length scales of ~15nm in diameter.

Single crystal PZN-8%PT (unpoled flux-grown single crystal, Microfine Materials Technologies Pte. Ltd., Singapore) was selected for study. Thin crystals were prepared along pseudocubic axes of $[100]_{PC}$, $[001]_{PC}$ and $[111]_{PC}$ using the method described previously [27]. The same sample preparation procedure was applied successfully for the determination of symmetry in single crystal $BaTiO_3$ [28].

The principle of domain identification is based on CBED determination of crystal symmetry, which has the spatial resolution ranging from few to hundreds of nanometer [29-31]. By scanning over a region of the crystal, ferroelectric domains can be identified by the change of CBED pattern symmetry (Fig. 1). For example, the mirror direction can be used to determine the 60° domains in PMN-31%PT with the aid of dynamic diffraction simulation using the Bloch wave method [25, 32]. To quantify the symmetries of the CBED patterns, normalized cross-correlation ($\gamma_m$) values of a pair of diffraction discs related by mirror symmetry are computed using the algorithm previously proposed by Kim *et al.* [33]. For convenience of having just one $\gamma_m$ value for one CBED pattern, the $\gamma_m$ values of three pairs of discs with the highest intensity were averaged, noted as $\gamma_{m,average}$ shown in Fig. 1(a). Spatial distribution of different CBED patterns are indicated by different colors in Fig. 1(d), which are used to represent the measured $\gamma_{m,average}$ [25]. The BF disk (transmitted beam) of CBED possesses the center of symmetry belonging to the Laue diffraction group according to Buxton *et al.* [29]. The location of the center of symmetry changes when crystal rotates as illustrated in Fig. 1(b) and (c).

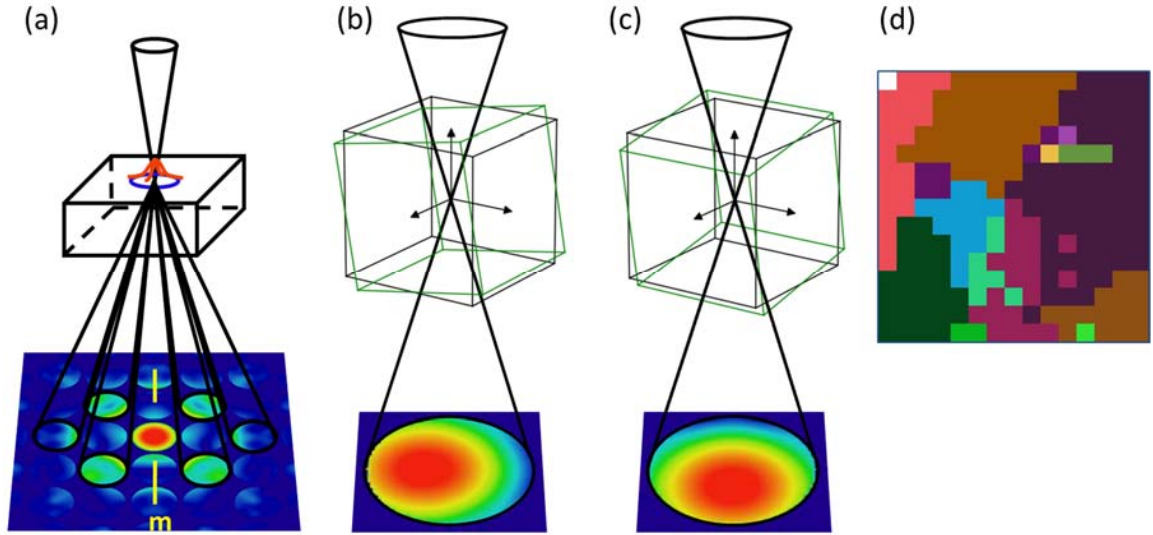

*Figure 1. Principles of using CBED for determining mirror symmetry and crystal rotation. Figure (a) shows an example for the mirror symmetry quantification, while crystal rotation along the x- and y-axes leads to a shift in the center of the CBED (000) pattern as shown in (b) and (c). The average of the cross-correlation coefficients of three pairs of discs in (a) is taken as $\gamma_{m,average}$, whose values are shown in (d) for a scan of 15x15 points or 225 CBED patterns. Here each color represents a different CBED pattern, whereas similar CBED patterns are shown in the same color.*

The SCBED experiments were carried out using a JEOL 2010F FEG TEM operated at 200kV with a convergent beam of 2.6nm in FWHM (full-width half-maximum). Energy-filtering, which improves the contrast of CBED patterns, was performed using a Gatan imaging filter (GIF). EF-SCBED was performed by scanning the focused electron probe over a selected area on a 15 x 15 grid, step size of 2nm, and through a post-column GIF energy window of 10eV. The shift and tilt of diffraction patterns during beam scanning were minimized and calibrated using a silicon single crystal [34]. Following the procedures described in [32], the symmetry of PZN-8%PT was determined as monoclinic *Pm*, which agrees with the X-ray diffraction result [12].

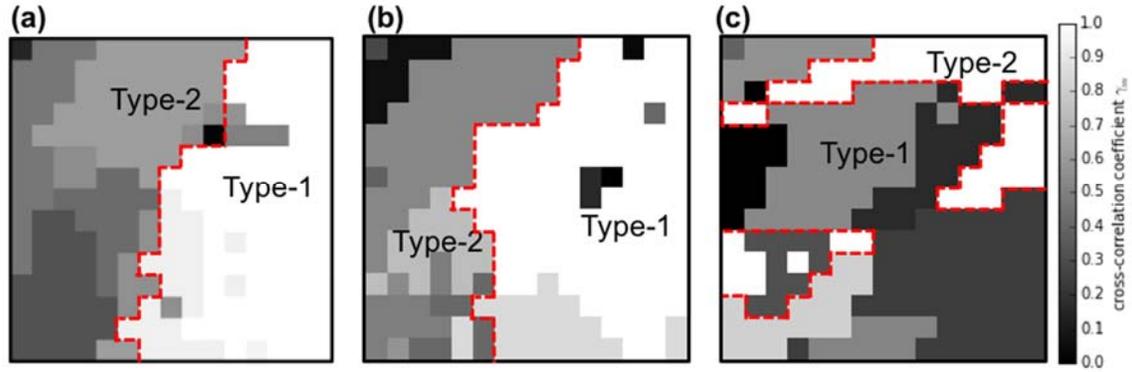

*Figure 2. Distribution of two nanodomains using SCBED. (a), (b) and (c) map out the $\gamma_{m,average}$ variations across two types of domains. The red dashed line indicates the domain boundary.*

Nanodomains are observed using EF-SCBED. Symmetry variations across these domains in three EF-SCBED datasets from three different sample areas are shown in Figs. 2(a), (b), and (c). The scan consists of 15 by 15 points, with a step size of 2 nm. The $\gamma_{m,average}$ of the representative CBED patterns in each region are shown in greyscale. We identified type-1 and type-2 nanodomains with different mirror symmetry. The boundaries between these two domains are indicated as dashed lines in the figures.

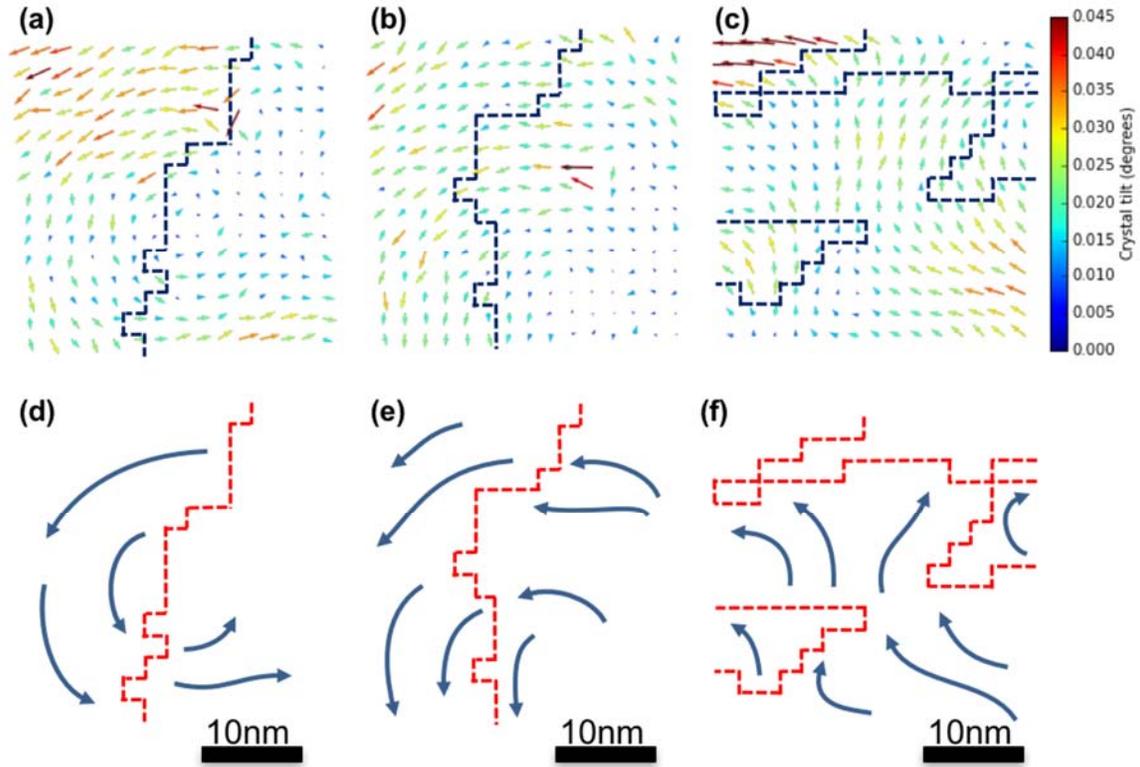

*Figure 3. Maps of distribution of two nanodomains and lattice rotation vortices. Figs. (a), (b), and (c) show the crystal rotation at each pixel, superimposed with the domain walls indicated by the blue dashed lines. Figs. (d), (e), and (f) illustrates how the crystal rotates across the domain boundaries schematically.*

We noticed that the center-of-mass of the intensity distribution within the BF disc of each pattern in the EF-SCBED dataset is not always located at the exact center. This observation could have two possible explanations: microscope optics and local crystal tilting. First, the hysteresis in the scanning coils or the lens in the microscope could lead to imperfect optical alignment while scanning the beam, which results in an effective beam-tilt and a consequent intensity redistribution in the BF disc. We excluded this effect by performing EF-SCBED on a Si single crystal. This measurement defines the maximum electron beam tilt and the lattice rotation measurement precision at $\pm 0.012$ degrees. Second, if the crystal is not oriented on the exact zone axis, this small angular deviation could also lead to an intensity redistribution in the BF disc. This is shown schematically in Figs. 1(b) and 1(c). In an effort to quantify how much the crystal is deviated from the exact zone axis, we calculated the displacements (in pixels) of the center-of-mass of each BF disc and converted these displacements into crystal rotations (in degrees).

By measuring the shift in the BF disc of a CBED pattern using this method, we determined the nanoscale rotation of the crystal and represented this rotation as a vector. The vector at each data point indicates the crystal rotation averaged over a volume of ~280nm$^3$. Figs. 3(a), (b), and (c) show the crystal rotation map derived from the same EF-SCBED datasets as Figs. 2(a), (b), and (c), respectively. Figure 3(a) shows a vortex-like pattern with the vortex center near the domain boundary, and a radius of curvature of ~7nm calculated from the discrete points. On the other hand, the vortex feature is not as distinct in Figs. 3(b) and (c). The continuous crystal rotation is shown schematically in Figs. 3(d), (e), and (f).

The type-1 and type-2 domains identified in Fig. 2 are associated with two distinguishable CBED patterns that were observed along the $[100]_{PC}$ incident direction (Figs. 4(a) and 4(b)). Figs. 2(a), (b), and (c) show the symmetry maps where these two patterns were detected. The highest $\gamma_m$ values of type-1 and type-2 patterns are detected along two different directions (A and B) as shown in Figs. 4(a) and 4(b). The A and B directions are rotated by 45° along the $[100]_{PC}$ zone axis. The corresponding simulated patterns for type-1 and type-2 domains are along monoclinic *Pm* zone axis $[100]_{Pm}$ and $[010]_{Pm}$, as shown in Figs. 4(c) and 4(d), respectively. In the *Pm* structure model, the polarization direction is $\vec{P}_S = [u,0,v]_{Pm} = [3,0,4]_{Pm}$, which lies in the mirror plane of *Pm* symmetry [12]. Along the $[100]_{Pm}$ incident direction, the $(001)/(00\bar{1})$ reflections are related by the mirror, which is parallel to the A direction in Fig. 4(a). This mirror is not observed along the $[010]_{Pm}$ incident direction. The projection of the polarization lies approximately on the $(101)/(\bar{1}0\bar{1})$ reflections, which is parallel to the B direction in Fig. 4(b). The highest mirror symmetry in this case is detected along direction B in the simulated pattern (Fig. 4(d)) with $\gamma_{m,simulated} = 60\%$.

Quantification of mirror symmetry for Fig. 4(a) and (b) gives $\gamma^1_{m,A} = 95\%$ and $\gamma^2_{m,A} = 34\%$, respectively (The superscript indicates the domain type, and the subscript denotes the mirror plane direction). This shows that the mirror plane of type-1 domains is along the A direction. For the type-2 domains, a good match is obtained with $[010]_{Pm}$. The $\gamma_m$ value along the B direction of the recorded patterns roughly agrees with the simulated value, with $\gamma_{m,experimental} = 54\%$.

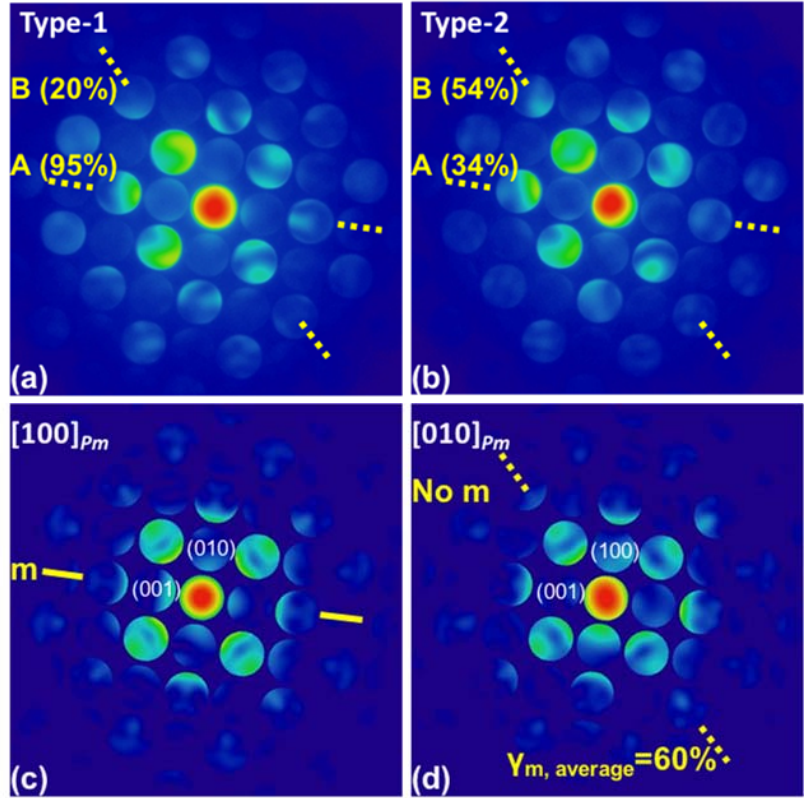

Figure 4. Experimental and simulated CBED patterns along various zone axes. *The mirror plane in the (a) type-1 and (b) type-2 domains is rotated by 45°. Figs. (c) & (d) show simulated patterns of $M_C$ (Pm) using the Bloch wave method and corresponding to the experimental (a) & (b) patterns, respectively. The indexing is based on simulated diffraction patterns.*

Based on the best matching structural model of *Pm*, the orientation relationship between the type-1 and 2 nanodomains with respect to the pseudocubic axes is shown schematically in

Figure 5. For type-1 domains, which belong to the $[100]_{Pm}$ zone axis, the monoclinic axes of $a_{Pm}$ and $b_{Pm}$ are along $[100]_{PC}$ and $[010]_{PC}$, respectively. The $c_{Pm}$ is slightly deviated away from the $[001]_{PC}$ direction with an angle $(90° − β)$ in the $a_{Pm} − c_{Pm}$ plane. Type-2 domains belong to the $[010]_{Pm}$ zone axis, for which the monoclinic axes of $a_{Pm}$ and $b_{Pm}$ are rotated by 90° with respect to the cubic c-axis. If converting the two polarization directions $[3,0,4]_{Pm}$ and $[0,3,4]_{Pm}$ from fractional coordinates into Cartesian coordinates, the polarization directions in Cartesian coordinates would be [3.03, 0.02, 4.05] and [0.02, 3.01, 4.05], respectively. The angle between the two vectors is 50°. The presence of 50° polarization domains is also evidence which excludes the T- or R-symmetries, since this type of domain is only permitted in crystals with orthorhombic or lower symmetries [35, 36].

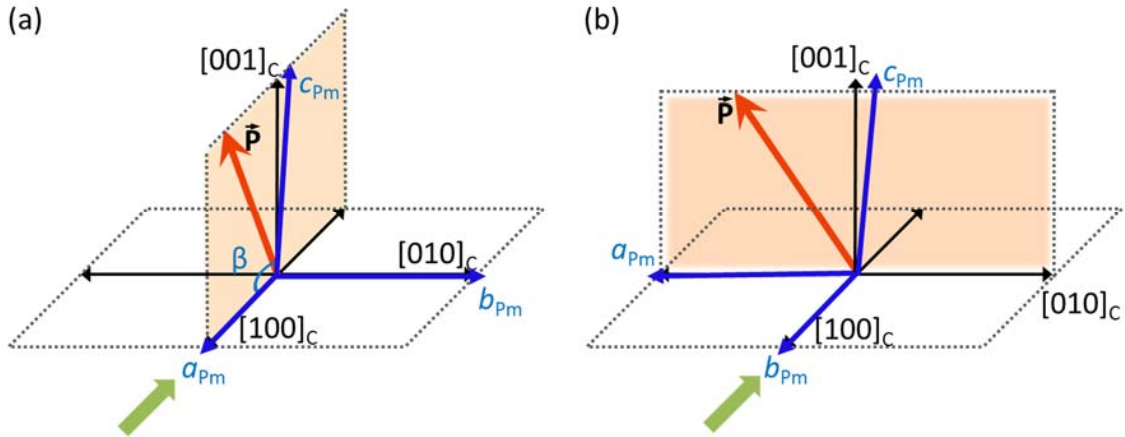

*Figure 5. Orientation relationship between two nanodomains with respect to the pseudocubic axes. Figs. (a) and (b) correspond to type-1 and type-2 domains, respectively.*

Flux-closure domain patterns associated with continuous dipole rotations have been reported in ferroelectric thin films [21, 22, 37-39] or ferroelectric nanodots [40-42] involving crystal systems with R or T symmetry. These patterns involve continuous dipole rotations near the vertices of triangular domain boundaries. The lattice rotation vortices or half-vortices observed near the boundary of two adjacent monoclinic domains involves small rotation of the crystal lattice having an average value of ~0.035 degrees.

The rotation we observed is part of the lattice deformation matrix with displacement vector $\boldsymbol{u}(\vec{r})$, defined by rigid body rotation tensor $\widetilde{w}_{ij} = 1/2\,(e_{ij} - e_{ji})$, where the strain tensor is $\varepsilon_{ij} = 1/2\,(e_{ij} + e_{ji})$ and the quantity $e_{ij} = \frac{\partial u_i}{\partial x_j}$. For relaxor ferroelectric crystals with monoclinic symmetry, disinclination exists between two domains with an angular mismatch determined by unit cell parameters [36, 43]. Strain accompanies the polarization rotation due to the strong electromechanical coupling [9, 10, 23]. We speculate that the crystal rotation vortex can be a result of accommodating disinclination strain and charge discontinuity. First, the disinclination strain can be estimated by calculating Lagrangian finite strain tensors [44]. Lattice parameters of two neighboring monoclinic *Pm* unit cells, distorted along two directions as depicted in Figure 5, are input parameters for calculating the strain tensors. The maximum strain at the domain wall is 1.3%, which is comparable to the 1.5% strain at the vertex core of rhombohedral $BiFeO_3$ [37]. Second, the charge discontinuity can be simplified by considering the surface charge density of a type-1 domain, a type-2 domain, and at the DW, denoted $\sigma_1$, $\sigma_2$, and $\sigma_{DW}$, respectively. The imbalance of $\sigma_1$, $\sigma_2$, and $\sigma_{DW}$ causes the dipoles to align with the electric field; however, the aligned dipoles change the surface charge and hence the electric field which causes further dipole alignment. The lattice rotation vortex can be the equilibrium state of the feedback process.

In conclusion, we observed local crystal rotation vortex at the 50° monoclinic domain boundary. The crystal rotation vortex is attributed to disinclination strain and depolarization field due to charge discontinuity across the domain walls. The above observation raises important questions about the roles of the lattice rotation vortex in domain switching in ferroelectric systems. Previously, first-principles calculations have predicted an intermediate state having a coexisting toroidal moment and out-of-plane-polarization in ferroelectric nanoparticles [45, 46]. The occurrence of lattice rotation vortices at the ferroelectric domain walls suggest the merging of 2D and 1D topological defects. An analogy can be made with the presence of magnetic vortices, known as skyrmions. The interplay between spin, orbital, charge, and strain degrees of freedom associated with skyrmions suggests a complex landscape of topological defects in ferroelectrics that may be explored for new applications and functionalities.

**Acknowledgments**

The authors express many thanks for the helpful discussion with Dr. Kyou-Hyun Kim and Prof. David A. Payne. The work is supported by DOE BES (Grant No. DEFG02-01ER45923). Electron microscopy experiments were carried out at the Center for Microanalysis of Materials at the Frederick Seitz Materials Research Laboratory of University of Illinois at Urbana-Champaign.